\documentclass[aps,prb,twocolumn,floats,showpacs,superscriptaddress,nofootinbib]{revtex4-1}

\usepackage{graphicx,epsfig}
\usepackage{times,bbm}
\usepackage{graphics,dcolumn,bm,float}
\usepackage{amssymb,amsmath, rotate,color}
\usepackage{mathtools}
\usepackage{booktabs}
\usepackage{tcolorbox}
\usepackage{adjustbox}
\usepackage[pagebackref=false,colorlinks,linkcolor=magenta,citecolor=cyan,urlcolor=magenta]{hyperref}

\begin{document}

\title{Planar Hall supercurrent and  $\delta\phi$-shift in the topological Josephson junction}
\author{Morteza Salehi}
\affiliation{Department of Physics, Bu-Ali Sina University, Hamadan, Iran}
\date{March 2024}

\begin{abstract}
 We theoretically investigate Josephson junctions comprising superconductors and ferromagnets on the surface of three-dimensional topological insulators. We use Bogoliubov-deGennes formalism and show the in-plane magnetization creates a difference between the upward and downward population of Andreev modes and produces a planar Hall supercurrent. Due to the strong spin-orbit interaction of Dirac fermions, bending on the supercurrent imposes a spin transfer torque on the junction. We develop a theory and demonstrate the relation between planar Hall supercurrent and spin transfer torque. The parallel component of in-plane magnetization creates an anomalous supercurrent that can flow even in zero superconducting phase difference and make $\delta\phi$-junction. We show in some range,$\pi/2d \leq m_y \leq \pi/d$, there is a $\pi$ shift in the Josephson supercurrent.  This research advances our understanding of quantum transport in 3DTIs and highlights their potential in emerging quantum technologies.
\end{abstract}

\maketitle
\section{Introduction}
Three-dimensional topological insulators (3DTIs) are a remarkable material class that has garnered considerable attention for their distinct electronic properties. The surface states of 3DTIs are host to two-dimensional Dirac fermions, uniquely characterized by their strong coupling of spin and momentum\cite{Fu2007PRL, Hasan2010RMP, Shen2013Book, Beenakker2015RMP}.
Spin transfer torque (STT) is the mechanism by which angular momentum is transferred between the electron's spin and the magnetization direction, driving intriguing phenomena such as the rotation of magnetization\cite{Tsymbal2019Book}. On the other hand, the planar Hall effect occurs when the direction of magnetic fields or magnetization lies perpendicular to the surface. These intricate interactions between electronic states, magnetic textures, and quantum topology make the exploration of quantum transport in 3DTIs a compelling and fertile ground for research, offering fundamental insights and promising applications in condensed matter physics\cite{Nakahara2003Book}.

A pivotal achievement has been the induction of superconductivity and ferromagnetism into the surface of 3DTIs through the proximity effect\cite{Bernevig2013Book,Qi2011rMP}. This breakthrough has laid the foundation for investigating novel quantum phenomena\cite{Linder2017BookSection, Hogl2015PRL, Eschrig2015RPP, Alicea2012RPP}, enabling the emergence of intriguing effects such as Majorana bound states\cite{Fu2008PRL}, Josephson supercurrent\cite{Josephson1962, Tanaka2009PRL, Pientka2017PRX}, and unconventional superconductivity\cite{Linder2010PRL, Linder2010PRB}.

\begin{figure}
\includegraphics*[scale=0.26]{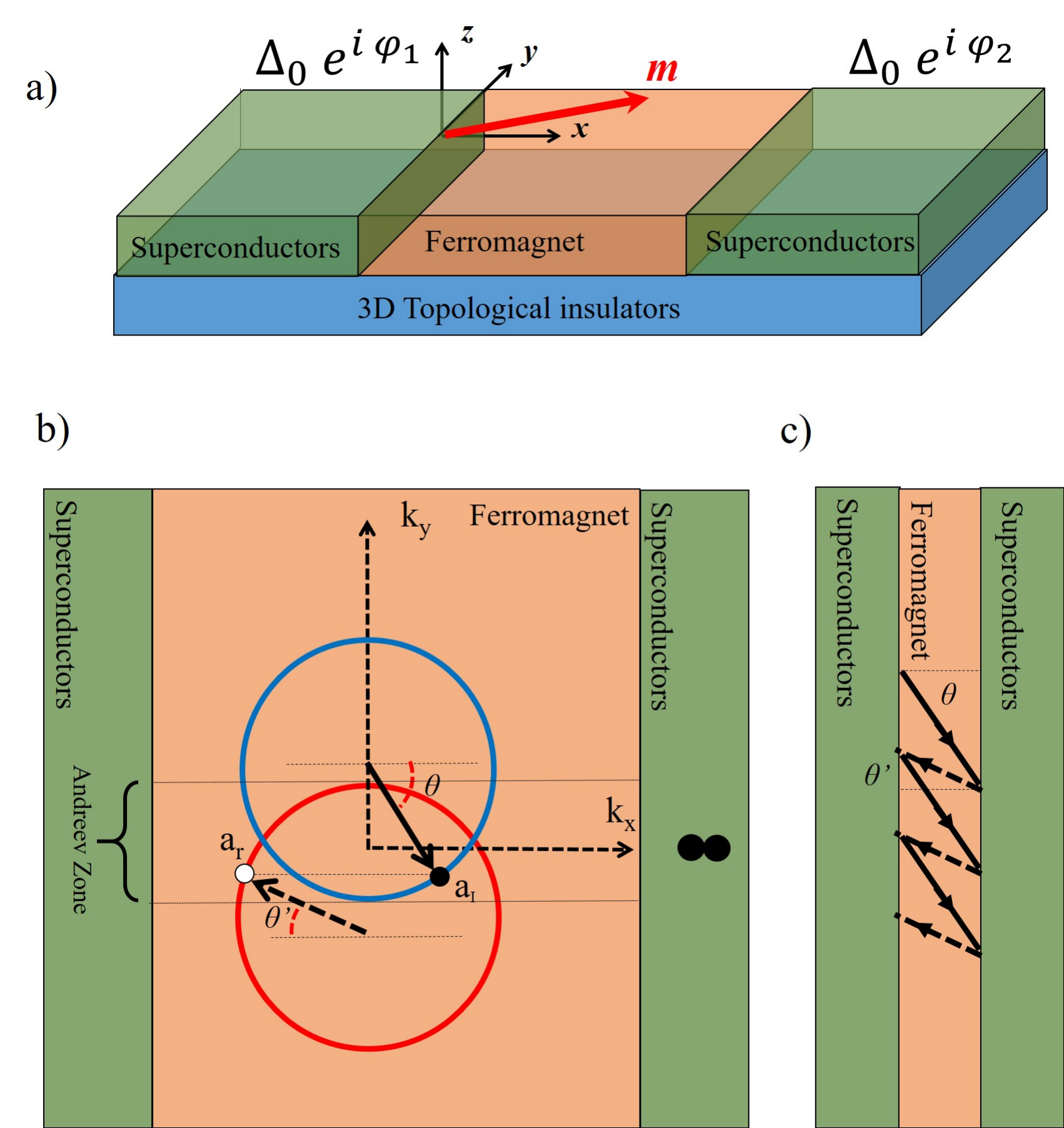}
\caption{(color online) a)The schematic illustration of Josephson junction on the surface of 3DTIs. In-plane magnetization, $\bm{m}$, lies in the $xy$-plane. b) The Andreev modes can be formed in the $k$-space. The blue circle relates to the electron-like quasi-particles, whereas the red one belongs to the hole-like excitations. In the presence of $m_x$ component, the Dirac cones separate in the $k_y$ direction equal to $2m_x$. c) In the real space, the Andreev modes propagate parallel to the interface.}
\label{Fig1.ToyModel}
\end{figure} 

In the following sections, we delve into the intriguing results of our theoretical study employing the Bogoliubov-deGennes (BdG) formalism. We explore a superconductor/ferromagnet/superconductor (SFS) junction with two distinct scenarios:

First, in-plane magnetization with a component perpendicular to the SFS junction's interface converts the Andreev-bound states into Andreev modes. This effect generates a planar Hall supercurrent (PHS) that flows parallel to the superconducting interfaces. We present these modes' energy-phase relation (EPR) and quantify the PHS's magnitude. Since Cooper pairs made by electric charges and they are a conserved quantity. The PHS causes a reduction in the Josephson supercurrent. The propagation direction of charge carrier changes by in-plane magnetization. The strong spin-orbit interaction results in any change in the carriers' propagation direction, leading to a modification in their spin texture. This phenomenon becomes evident through a torque exerted on the junction\cite{Salehi2021}. We developed a theory to show the relationship between STT and PHS. The dynamics of the spin density wave lead to an exerted torque on magnetization direction and illustrate the relationship between quantum topology and magnetic order. These findings provide crucial insights into the emergent phenomena in quantum transport, with potential applications for future technology.

Second, in-plane magnetization with a parallel component with respect to the junction's interface, resulting in an anomalous supercurrent, the $\delta\phi$-junction. In zero phase difference, a non-zero supercurrent flows through the SFS junction \cite{Andreev1991, Sigrist1998PTP, Ryazanov2001PRL, Braude2007PRL, Houzet2007PRB, Buzdin2008PRL, Shukrinov2022PU, Goldobin2011PRL, Goldobin2015PRB, Szombati2016NP}. This phenomenon hinges on the interplay between the unique EPR of Andreev modes and ferromagnetism. Moreover, it can lead to the emergence of $\pi$-junctions, where the supercurrent changes sign upon a $\pi$ phase difference. We delve into the physics of $\delta\phi_0$ junctions and the underlying principles that govern their behavior.


\section{Theory and formulation}
At the interface of superconductors, an incident electron from the non-superconducting side can undergo Andreev reflection, where it is reflected back as a hole while a Cooper pair passes into the superconducting lead \cite{Andreev1964JETP}. This phenomenon, dominant at voltages lower than the superconducting gap, transforms the dissipative current into a dissipationless one\cite{BTK}.
In part (a) of Fig.(\ref{Fig1.ToyModel}), we examine a SFS junction on the 3DTI, assuming that both magnetization and superconductivity can be induced separately by means of the proximity effect. The in-plane magnetization, $\bm{m}=(m_x, m_y)$, lies in the $xy$-plane and is confined to $0 \leq x \leq L$, while the s-wave superconductivity covers $-\infty \leq x \leq 0$ and $L\leq x \leq +\infty$. In part (b) of the diagram shown in Fig. (\ref{Fig1.ToyModel}), the junction is depicted in the $k$-space. The $x$-component of the magnetization causes a relative separation of the Dirac cones in the direction of $k_y$. The blue ring represents unoccupied states within the electron-like cone, with energy $\epsilon$ exceeding the chemical potential $\mu$. The radius of this electron-like circle corresponds to $\mu + \epsilon$. The presence of a black dot, denoted as ($a_I$), signifies an incoming fermion. Additionally, the black arrow, emanating from the center of the blue circle, indicates the propagation direction of the incoming fermion in real space, determined by its group velocity.

In the ballistic regime, the energy and parallel component of the wave vector remain constant during scattering processes at the junction's interfaces. As a result, the radius of the hole-like ring (the red one) becomes $|\mu-\epsilon|$. The Andreev zone is defined by the overlap of the electron-like and hole-like rings. There is a probability of $R_A$ for the incoming fermion to find an empty state on the hole-like ring. The conservation of the $k_y$ component determines the hole's location on the red ring. Furthermore, the hole's position on the valence or conduction bands specifies whether it follows a specular or retro direction\cite{Beenakker2006PRL}. This effect is illustrated in part (b) of Fig.(\ref{Fig1.ToyModel}).

A similar scenario can occur for the reflected hole on the other interface of the Josephson junction. The hole can be reflected as a fermion in the $a_I$ location of the electron-like ring, creating an Andreev mode at the junction that can be propagated parallel to the superconducting interfaces. The propagation direction of each state in real space is indicated by a black arrow centered at the origin of the related ring. This effect is schematically depicted in part (c) of Fig.(\ref{Fig1.ToyModel}). These Andreev modes create the supercurrent that flows between the superconductor leads\cite{deGennes1966}. Additionally, the incident fermion has the probability of $R_N$ for normal reflection. The conservation of probabilities ensures that $R_N + R_A = 1$. Outside the Andreev zone, incoming fermions undergo perfect normal reflection ($R_N = 1$), leading to anisotropic angle-dependent Andreev reflection\cite{Salehi2023}. The different directions of the incoming fermion and its Andreev hole result in a PHS, detectable via a four-terminal setup. Its physics is rooted in the propagation bending of charge carriers, which, due to full spin-orbit interaction on the surface of 3DTIs, modifies the spin configuration of the carriers and leads to torque imposed on the interface and magnetization direction.
Furthermore, for incoming fermions with energy levels exceeding $\Delta_0$, alternative possibilities are available to them, such as filling the stationary states far above the superconducting gap.

On the other hand, the $y$-component of magnetization displaces the electron-like and hole-like rings within the $k_x$ direction of the $ k$-space, adding an extra phase in charge carriers' wave functions, resulting in a $\delta\phi$-junction
\subsection{Model}
To investigate the junction theoretically, we employ the Bogoliubov-deGennes (BdG) equation\cite{deGennes1966} given by

\begin{equation}
\begin{array}{ll}
H_{BdG}= & \hbar v_F \eta_z \otimes \left((\boldsymbol{\sigma}\times \boldsymbol{k}).\hat{e}_z\right)-\eta_0 \otimes (\textbf{m}.\boldsymbol{\sigma}) \\
& \\
& +\Delta_0 (\eta_x \otimes \sigma_0 \cos\phi+\eta_y \otimes \sigma_0 \sin\phi) \\
\end{array}.
\label{Eq.HBdG}
\end{equation}
The first term describes the two dimensional Dirac Hamiltonian. The second term comes from the Stoner model and $\textbf{m}$ illustrates the role of magnetization. The superconductivity enters with the $\Delta_0$ in Eq.(\ref{Eq.HBdG}). Here $\boldsymbol{\sigma}$ and $\boldsymbol{\eta}$ represent the Pauli vector in spin and Nambu spaces, respectively. Also, $v_F$ denotes the Fermi velocity. We set $\hbar v_F=1$ for simplicity\cite{Salehi2021}. 

In the magnetic region, the eigenvalues are given by

\begin{equation}
\epsilon_{\pm}=\pm \sqrt{(k_x\pm m_y)^2+(k_y\mp m_x)^2}\pm \mu,
\label{Eq.Dispersion1}
\end{equation}

where the $\epsilon_\pm$ stands for electron and hole states respectively. Eq.(\ref{Eq.Dispersion1}) describes two cones in $k$-space. Without magnetization, the cones are located at the origin. In-plane magnetization, $\{m_x \neq 0, m_y\neq 0\}$, adjusts their locations and separates them from each other by $2\sqrt{m_x^2+m_y^2}$\cite{Yokoyama2010PRB}. 

The corresponding states  of electrons can be obtained as \cite{Salehi2023}

\begin{equation}
\Phi^\pm_e(r)=\left(\begin{array}{c}
i \\
\pm e^{\pm i \theta_e}\\
0\\
0\\
\end{array}\right)e^{\pm i k^e_x-im_yx+ik_yy}.
\label{Eq.Psie}
\end{equation}

where the $\pm$ sign refers to the right-going or left-going states, respectively. The hole-like wave function corresponding to the eigenvalues of Eq.(\ref{Eq.Dispersion1}) can be derived as

\begin{equation}
\Phi^\pm_h(r)=\left(\begin{array}{c}
0 \\
0\\
i\\
\pm \mathcal{S} e^{\pm i \theta_h}\\
\end{array}\right)e^{\pm i \mathcal{S}k^h_x x+im_yx+ik_y y}.
\label{Eq.psih}
\end{equation}

Here, $\mathcal{S}=sign(\epsilon-\mu)$ tunes the reflected hole in the conduction or valence band. The $\theta_e$ and $\theta_h$ stand for the direction of propagation,

\begin{equation}
     \begin{array}{cc}
        \theta_e=\tan^{-1}\left(\frac{k_y-m_x}{k^e_x+m_y}\right) , &  \theta_h=\tan^{-1}\left(\frac{k_y+m_x}{k^h_x-m_y}\right) \\
          
     \end{array}.
     \label{Eq.thetas}
\end{equation}

Using $k_y$ conservation in Eq.(\ref{Eq.Dispersion1}), the reflected holes find a stable state in the Andreev zone if they satisfy the condition $  |\epsilon-\mu|\geq |k_y+m_x| $. From an experimental perspective, the induction of superconductivity on the surface of 3DTIs requires a sufficient density of states. To satisfy this condition, we utilize a heavily doped approximation, $\mu_S \rightarrow \infty$\cite{Beenakker2006PRL, Titov2006PRB, Zhang2008PRL}. The wave functions in the presence of superconductivity, $\Delta_0 \neq 0$, can be derived in a similar way,

\begin{equation}
     \begin{array}{l}
    \psi^{S,\pm}_e(\textbf{r})=\left( \begin{array}{c}
         i  \\
         \pm1 \\
         i e^{-i\phi}e^{-i\beta}\\
         \pm e^{-i\phi}e^{-i\beta}\\
    \end{array} \right)e^{\pm ik^{S,e}_x x+i k_y y} \\
    \\
    
        \psi^{S,\pm}_h(\textbf{r})=\left( \begin{array}{c}
         i e^{i\phi}e^{-i\beta} \\
         \mp e^{i\phi}e^{-i\beta} \\
        i\\
         \mp 1\\
    \end{array} \right)e^{\pm ik^{S,h}_x x+i k_y y}   
     \end{array}.
\end{equation}

Here, $\phi$ is the phase of the superconducting gap. Since it can be different in the left and right leads, we define the superconducting phase difference as $\delta\phi=\phi_L-\phi_R$. Moreover, we define $\beta=\cos^{-1}(\epsilon/\Delta_0)$. The $x$-component of the wave vector in the superconducting region can be derived as

\begin{equation}
 \begin{array}{l}
    k^{S,e}_{x}= \sqrt{(\mu_S+\sqrt{\epsilon^2-\Delta_0^2})^2-k_y^2}  \\
    \\
    k^{S,h}_{x}=  \sqrt{(\mu_S-\sqrt{\epsilon^2-\Delta_0^2})^2-k_y^2} ,
 \end{array}
 \label{Eq.KSh}
 \end{equation}

where the index of $S$ refers to the superconducting region.

\subsection{Energy-phase  relation of Andreev modes}

To obtain the EPR of the Andreev modes, we construct the whole wave functions of each region as below,
\begin{equation}
    \begin{array}{ll}
    \Psi^S_L(\textbf{r})=a_1 \psi^{S,-}_e(\textbf{r})+a_2\psi^{S,-}_h(\textbf{r}), \\
    \\
    \Psi^{F}(\textbf{r})=a_3\psi_e^{+}(\textbf{r})+a_4\psi_e^{-}(\textbf{r})+a_5\psi_h^{+}(\textbf{r})+a_6\psi_h^{-}(\textbf{r}),\\
    \\
    \Psi^S_R(\textbf{r})=a_7 \psi^{S,+}_e(\textbf{r})+a_8\psi^{S,+}_h(\textbf{r}). \\
    \end{array}
\end{equation}

The $L$ and $R$ belong the left and right superconducting electrodes. Each amplitude of $\{a_1, ..., a_8\}$ relates to its state. If we set $d$ as the width of the junction, the boundary conditions that match the whole wave functions of each region with its neighbor are,
\begin{equation}
    \begin{array}{cc}
         \Psi^S_L(x=0)=\Psi^F(x=0), \\
         \\
         \Psi^F(x=d)=\Psi^S_R(x=d) .
    \end{array}
    \label{Eq.BoundaryCondition}
\end{equation}
The Eq.(\ref{Eq.BoundaryCondition}), leads to eight homogeneous equations for obtaining the $\{a_1, ..., a_8\}$ amplitudes. To have a non-trivial solution, the determinant of its coefficient's matrix must be zero. After some straightforward calculations, one can reach the below equation,

\begin{equation}
\mathcal{A}\cos(\delta\phi+2m_yd)+\mathcal{D}=\mathcal{B}\cos(2\beta)+\mathcal{C}\sin(2\beta).
\label{Eq.energy1}
\end{equation}
Here we have,
\begin{equation}
    \begin{array}{l}
    \mathcal{A}=\cos(\theta_e)\cos(\theta_h), \\
    \\
    \mathcal{B}=\cos(k_ed)\cos(k_hd)\cos(\theta_e)\cos(\theta_h)-\sin(k_ed)\sin(k_hd),\\
    \\
    \mathcal{C}=\cos(k_ed)\sin(k_hd)\cos(\theta_e)+\sin(k_ed)\cos(k_hd)\cos(\theta_h),\\
    \\
    \mathcal{D}=\sin(ke_d)\sin(k_hd)\sin(\theta_e)\sin(\theta_h).
    \end{array}
\end{equation}
To proceed further, we use the experimental point of view and take the length $d$ of the junction is much less than the superconducting coherence length, $\hat{\xi}=\hbar v_F/\Delta_0$\cite{Titov2006PRB, Tanaka2009PRL}. In this limit, known as \textit{short-junction limit}, one can neglect $\epsilon$ compared to $\mu$. Finally, the EPR of the Andreev mode can be derived from Eq.(\ref{Eq.energy1}) as,

\begin{equation}
    \epsilon(\delta\phi)=\Delta_0 \cos\left(
    \begin{array}{l}
\frac{1}{2}\cos^{-1}\left(\frac{\mathcal{D}+\mathcal{A}\cos(\delta\phi+2m_yd)}{\sqrt{\mathcal{B}^2+\mathcal{C}^2}}\right)
    \\
    \\
    +\frac{1}{2}\cos^{-1}\left(\frac{\mathcal{B}}{\sqrt{\mathcal{B}^2+\mathcal{C}^2}}\right) 
   \end{array}\right).
   \label{Eq.E-phase}
\end{equation}    

From Eq.(\ref{Eq.thetas}) and Eq.(\ref{Eq.Dispersion1}), we define the boundaries of the Andreev zone as below,
\begin{equation}
    \begin{array}{l}
    \theta_{max}=\sin^{-1}(\min\{1, 1-\frac{2m_x}{\mu}\})\\
    \\
    \theta_{min}=\sin^{-1}(\max\{-1, -1-\frac{2m_x}{\mu}\}). 
     \end{array}
     \label{Eq.ThetaCritical}
\end{equation}

In the Andreev zone, the EPR of Eq.(\ref{Eq.E-phase}) is real. In the next section, we obtain the Josephson current and the PHS.

\subsection{Spin Transfer Torque}
The surface states of 3DTIs have a strong coupling between their spin and momentum, with the dynamics of the spin-density wave serving as a valuable quantity for characterizing this interaction. We initiate our analysis with the second quantization form of Eq.(\ref{Eq.HBdG}) given by

\begin{equation}
\mathcal{H}_{BdG}=\int d^2r \hat{\Psi}^\dagger(\textbf{r})H_{BdG}\hat{\Psi}(\textbf{r}),
\label{Eq.(01)}
\end{equation}

where $\hat{\Psi}(\textbf{r})=(\hat{\Phi}_e(\textbf{r}), \hat{\Phi}_h(r))^T$, and $\hat{\Psi}^\dagger(\textbf{r})$ represent annihilation and creation Bogoliubov field operators. Here, $\hat{\Phi}_e(\textbf{r})$ acts as the annihilation field operator for electrons, while $\hat{\Phi}_h(\textbf{r})$ fulfills the same role for holes. Field operators obey the anti-commutation relations. The $z$-component of the spin-density wave is defined as

\begin{equation}
S_z=\hat{\Psi}^\dagger(\textbf{r}) (\eta_0\otimes \sigma_z) \hat{\Psi}(\textbf{r}),
\end{equation}

Utilizing the definition of Bogoliubov field operators, the spin-density wave can be partitioned into electron and hole components, leading to the expression

\begin{equation}
S_z=S_z^e+S_z^h,
\end{equation}

signifying that the dynamics of a spin-density wave can be viewed as the sum of its electron-like and hole-like components, as described by

\begin{equation}
\frac{\partial S_z}{\partial t}=\frac{\partial S^e_z}{\partial t}+\frac{\partial S^h_z}{\partial t}.
\end{equation}

Subsequently, we obtain

\begin{equation}
\begin{array}{l}
\frac{\partial S^e_z}{\partial t}=\frac{\partial \hat{\Phi}_e^\dagger(\textbf{r})}{\partial t}\sigma_z \hat{\Phi}_e(\textbf{r})+\hat{\Phi}_e^\dagger(\textbf{r})\sigma_z\frac{\partial \hat{\Phi}_e(\textbf{r})}{\partial t},\\
\\
\frac{\partial S^h_z}{\partial t}=\frac{\partial \hat{\Phi}_h^\dagger(\textbf{r})}{\partial t}\sigma_z \hat{\Phi}_h(\textbf{r})+\hat{\Phi}_h^\dagger(\textbf{r})\sigma_z\frac{\partial \hat{\Phi}_h(\textbf{r})}{\partial t}.
\end{array}
\end{equation}

The equation of motion can be employed to express the time derivatives of the electron-like and hole-like spin-density components as

\begin{equation}
\begin{array}{l}
\frac{\partial S^e_z}{\partial t}=i \left[ \mathcal{H}_{BdG}, \hat{\Phi}_e^\dagger(\textbf{r})\right]\sigma_z \hat{\Phi}_e(\textbf{r})+i \hat{\Phi}_e^\dagger(\textbf{r})\sigma_z \left[ \mathcal{H}_{BdG}, \hat{\Phi}_e(\textbf{r})\right],\\
\\
\frac{\partial S^h_z}{\partial t}=i \left[ \mathcal{H}_{BdG}, \hat{\Phi}_h^\dagger(\textbf{r})\right]\sigma_z \hat{\Phi}_h(\textbf{r})+ i \hat{\Phi}_h^\dagger(\textbf{r})\sigma_z \left[ \mathcal{H}_{BdG}, \hat{\Phi}_h^\dagger(\textbf{r})\right].
\end{array}
\label{Eq.14}
\end{equation}

The commutation relations associated with Eq.(\ref{Eq.(01)}) can be expressed as

\begin{equation}
\begin{array}{ll}
\left[\mathcal{H}_{BdG}, \hat{\Phi}_e(\textbf{r'}) \right]= & \left(-H_D(\textbf{k})+\textbf{m}.\boldsymbol{\sigma} \right) \hat{\Phi}_e(\textbf{r})-\sigma_0 \Delta_0 e^{i\phi} \hat{\Phi}_h(\textbf{r}),\\
\\
\left[\mathcal{H}_{BdG}, \hat{\Phi}_e^\dagger(\textbf{r'}) \right]= & \hat{\Phi}_e^\dagger(\textbf{r}) \left(H_D(-\textbf{k})-\textbf{m}.\boldsymbol{\sigma} \right) +\hat{\Phi}_h^\dagger(\textbf{r}) \sigma_0 \Delta_0 e^{-i\phi},\\
\\
\left[\mathcal{H}_{BdG}, \hat{\Phi}_h(\textbf{r'}) \right]= & \left(H_D(-\textbf{k})-\textbf{m}.\boldsymbol{\sigma} \right) \hat{\Phi}_h(\textbf{r})-\sigma_0 \Delta_0 e^{-i\phi} \hat{\Phi}_e(\textbf{r}),\\
\\
\left[\mathcal{H}_{BdG}, \hat{\Phi}_e^\dagger(\textbf{r}') \right]= & \hat{\Phi}_e^\dagger(\textbf{r}) \left(H_D(\textbf{k})+\textbf{m}.\boldsymbol{\sigma} \right) +\hat{\Phi}_e^\dagger(\textbf{r}) \sigma_0 \Delta_0 e^{i\phi}.
\end{array}
\label{Eq.CommutationRelation}
\end{equation}

Subsequently, we obtain

\begin{equation}
\begin{array}{ll}
\frac{\partial S^e_z}{\partial t}= & i \left\lbrace
\begin{array}{c}
\hat{\Phi}_e^\dagger(\textbf{r})H_D(-\textbf{k})\sigma_z\hat{\Phi}_e(\textbf{r})- \hat{\Phi}_e^\dagger(\textbf{r})\sigma_z H_D(\textbf{k})\hat{\Phi}_e(\textbf{r}),\\
\\
-\hat{\Phi}_e^\dagger(\textbf{r})\textbf{m}.\boldsymbol{\sigma}\sigma_z\hat{\Phi}_e(\textbf{r})+ \hat{\Phi}_e^\dagger(\textbf{r})\sigma_z\textbf{m}.\boldsymbol{\sigma}\hat{\Phi}_e(\textbf{r}),\\
\\
+\hat{\Phi}_e^\dagger(\textbf{r})\sigma_0 \Delta_0 e^{-i\phi} \sigma_z \hat{\Phi}_e(\textbf{r})-\hat{\Phi}_e^\dagger(\textbf{r})\sigma_z \sigma_0 \Delta_0 e^{i\phi}\hat{\Phi}_e(\textbf{r})
\end{array}
\right\rbrace
\end{array}.
\label{Eq.16}
\end{equation}

The hole-like spin-density wave can be derived similarly,

\begin{equation}
\begin{array}{ll}
\frac{\partial S^h_z}{\partial t}= & i \left\lbrace
\begin{array}{l}
\hat{\Phi}_e^\dagger(\textbf{r})H_D(\textbf{k})\sigma_z\hat{\Phi}_e(\textbf{r})- \hat{\Phi}_e^\dagger(\textbf{r})\sigma_z H_D(-\textbf{k})\hat{\Phi}_e(\textbf{r})\\
\\
+\hat{\Phi}_e^\dagger(\textbf{r})\textbf{m}.\boldsymbol{\sigma}\sigma_z\hat{\Phi}_e(\textbf{r})- \hat{\Phi}_e^\dagger(\textbf{r})\sigma_z\textbf{m}.\boldsymbol{\sigma}\hat{\Phi}_e(\textbf{r})\\
\\
+\hat{\Phi}_e^\dagger(\textbf{r})\sigma_0 \Delta_0 e^{i\phi} \sigma_z \hat{\Phi}_e(\textbf{r})-\hat{\Phi}_e^\dagger(\textbf{r})\sigma_z \sigma_0 \Delta_0 e^{-i\phi} \hat{\Phi}_e(\textbf{r})
\end{array}
\right\rbrace
\end{array}
\label{Eq.17}.
\end{equation}

The third row of Eq.(\ref{Eq.16}) and Eq.(\ref{Eq.17}) nullify each other, indicating a cancellation of terms. The quasi-particle group velocity operator, derived from Eq.(\ref{Eq.HBdG}), is expressed as

\begin{equation}
\vec{\textbf{V}}=\hat{V}_x \hat{e}_x+\hat{V}_y\hat{e}_y=-\eta_0\sigma_y \hat{e}_x+\eta_0\sigma_x \hat{e}_y
\label{Eq.(18)}.
\end{equation}

Due to its structure, the group velocity can be decomposed into electron-like and hole-like sections, defined by

\begin{equation}
\begin{array}{l}
\hat{\textbf{V}}^e=\hat{V}_x^e \hat{e}_x+V_y^e \hat{e}_y,\\
\hat{\textbf{V}}^h=\hat{V}_x^h \hat{e}_x+V_y^h \hat{e}_y.
\end{array}
\label{Eq.20}
\end{equation}

Here, $\hat{V}_x^e=\hat{V}_x^h=-\sigma_y$ and $\hat{V}_y^e=\hat{V}_y^h=\sigma_x$. Utilizing Eq.(\ref{Eq.20}), Eq.(\ref{Eq.16}) can be reformulated as

\begin{equation}
\begin{array}{ll}
\frac{\partial S^e_z}{\partial t}= & i \left\lbrace
\begin{array}{l}
(\boldsymbol{\nabla} \hat{\Phi}_e^\dagger(\textbf{r}))\times(\hat{\textbf{V}}^e\hat{\Phi}_e(\textbf{r})).\hat{e}_z,\\
+(\hat{\Phi}_e^\dagger(\textbf{r})\hat{\textbf{V}}^e)\times(\boldsymbol{\nabla} \hat{\Phi}_e(\textbf{r})).\hat{e}_z,\\
+2\hat{\Phi}_e^\dagger(\textbf{r})(\boldsymbol{\sigma}\times \textbf{m}).\hat{e}_z\hat{\Phi}_e(\textbf{r}),
\end{array}
\right\rbrace
\end{array}.
\label{Eq.16}
\end{equation}

Further simplification yields

\begin{equation}
\begin{array}{ll}
\frac{\partial S^e_z}{\partial t}= & i \left\lbrace
\begin{array}{l}
  \boldsymbol{\nabla}\times (\Phi^\dagger(\textbf{r})\hat{\textbf{V}}^e\Phi(\textbf{r})).\hat{e}_z,\\
+2\hat{\Phi}_e^\dagger(\textbf{r})(\boldsymbol{\sigma}\times \textbf{m}).\hat{e}_z\hat{\Phi}_e(\textbf{r}),
\end{array}
\right\rbrace
\end{array}.
\label{Eq.Se3}
\end{equation}

Similarly, the hole-like spin-density wave can be derived. It is evident that $\vec{\textbf{J}}=\vec{\textbf{J}}^e+\vec{\textbf{J}^h}$, where the electron-like and hole-like current densities are defined as

\begin{equation}
\begin{array}{l}
\vec{\textbf{J}}^e=\hat{\Phi}_e^\dagger(\textbf{r})\vec{\boldsymbol{V}^e}\hat{\Phi}_e(\textbf{r}),\\
\vec{\textbf{J}}^h=\hat{\Phi}_e^\dagger(\textbf{r})\vec{\boldsymbol{V}^h}\hat{\Phi}_e(\textbf{r}).
\end{array}
\label{Eq.CurrentDensityDefinition}
\end{equation}

By combining the electron and hole sections, we arrive at

\begin{equation}
\frac{\partial S_z}{\partial t}=  i \left\lbrace
\begin{array}{l}
  (\boldsymbol{\nabla}\times \boldsymbol{J}).\hat{e}_z,\\
+2\hat{\Psi}^\dagger(\textbf{r})(\boldsymbol{\sigma}\times \textbf{m}).\hat{e}_z\hat{\Psi}(\textbf{r}),
\end{array}\right\rbrace
\label{Eq.PartialS}
\end{equation}

Due to the non-curvature feature of $\hat{e}_z$, we can rearrange Eq.(\ref{Eq.PartialS}) to obtain a continuity equation

\begin{equation}
\frac{\partial S_z}{\partial t}+i \boldsymbol{\nabla}.( \hat{e}_z \times \hat{\boldsymbol{J}})=\hat{\Psi}^\dagger(\textbf{r})2(\boldsymbol{\sigma}\times \textbf{m}).\hat{e}_z\hat{\boldsymbol{\Psi}}(\textbf{r})
\label{Eq.SpinDensity}
\end{equation}

As the current is a conserved quantity, the second term on the left side of Eq.(\ref{Eq.SpinDensity}) represents the current bending, acting as the $z$-component of spin-density current, denoted as $J^s_z=\hat{e}_z \times \hat{\boldsymbol{J}}$. Furthermore, the source of Eq.(\ref{Eq.SpinDensity}) is the density of STT, given by $dT_z=\hat{\Psi}^\dagger(\textbf{r})2(\boldsymbol{\sigma}\times \textbf{m}).\hat{e}_z\hat{\boldsymbol{\Psi}}(\textbf{r})$.

Because of strong spin-orbit interaction the spin of electrons and hols lie in the xy-plane. So, the $S_z$ and its time derivation, $\partial S_z/\partial t=0$, are zero. The integration of Eq.(\ref{Eq.SpinDensity}) over the SFS junction results to the $z$-component of STT as

\begin{equation}
T_z= \int dT_z d^2r= \int \boldsymbol{\nabla}.\textbf{J}^s_z d^2 r.
\label{Eq.CTT1}
\end{equation}

The divergence theorem leads to an integral over the closed loop as
\begin{equation}
T_z=\oint J^s_z.d\textbf{l}
\label{Eq.CTT2}.
\end{equation}

Eq.(\ref{Eq.CTT2}) shows any reduction of the current in one direction compensates by an increase in the other. So, it means the STT is directly related to the PHS that flows parallel to the interface.


\section{Results and discussion}
\subsection{The energy-phase diagram}
\begin{figure}
\includegraphics*[scale=0.30]{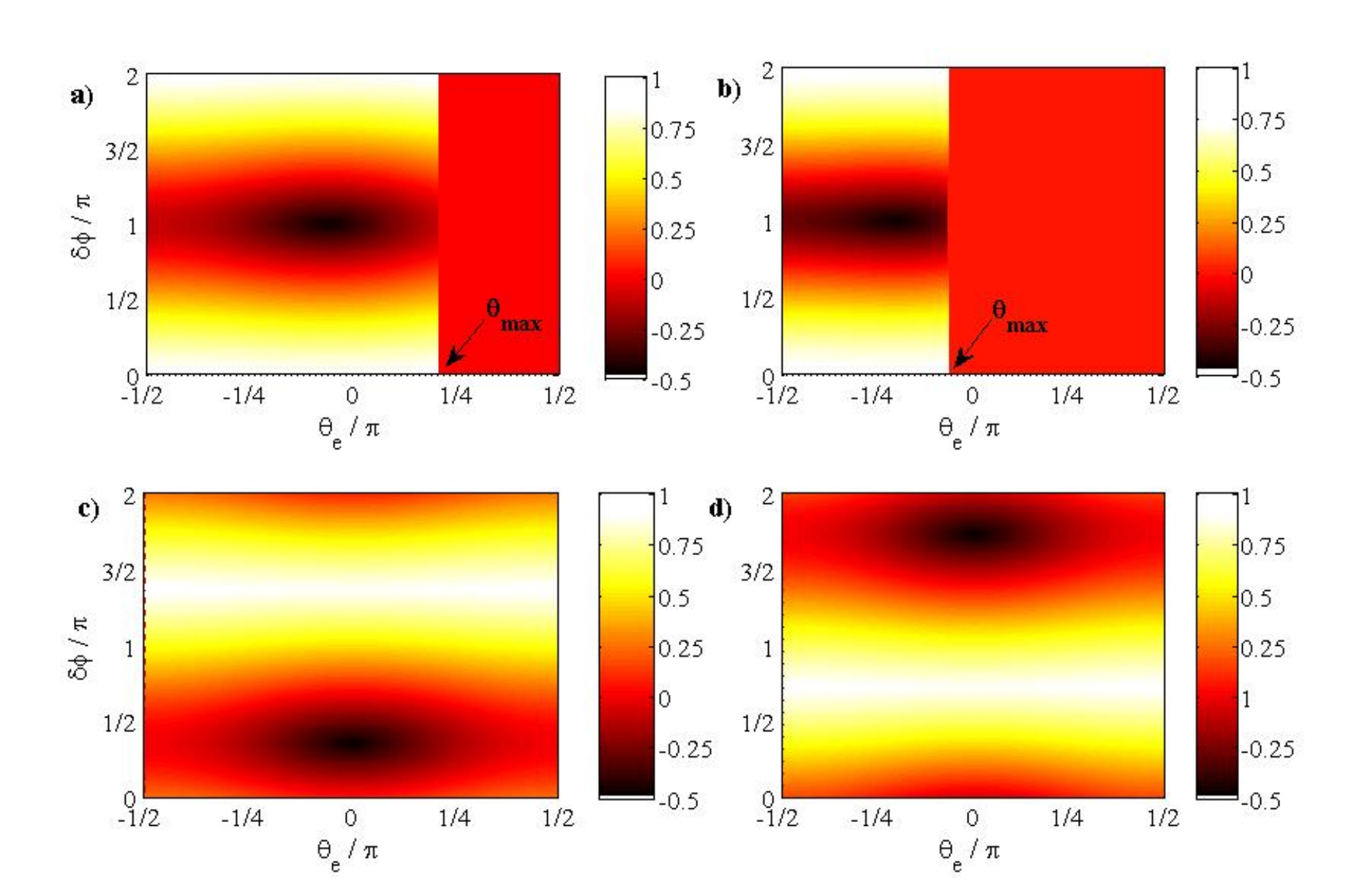}
\caption{(Color online) The energy-phase relation (EPR) of Andreev modes versus propagating angle of incoming fermions and superconducting phase difference. At a) and b), we set $m_x=0.2 \Delta_0$ and $m_x=0.6 \Delta_0$, respectively, whereas $m_y=0$. The $\theta_{max}$ shows the critical angles for creating Andreev modes. For c) and d), we set $m_y=\Delta_0$ and $m_y=2\Delta_0$, respectively, whereas $m_x=0$. In this case, the movement of Dirac cones occurs in the $ k_x$ direction. The width of the junction is $d=\hbar v_F/\Delta_0$.}
\label{Fig1.Energy-phase}
\end{figure} 
The Eq.(\ref{Eq.E-phase}) describes the EPR of Andreev modes. In the presence of $m_x$,  the Fermi circles of Fig.(\ref{Fig1.ToyModel}) separate in $k_y$-direction. The overlap of these circles creates the Andreev zone. In this zone, the Andreev modes are stable, and their energy is real. Its boundaries are entirely determined by Eq.(\ref{Eq.ThetaCritical}). Outside of this zone, the Andreev modes are unstable with complex EPR.  In the short-junction limit, there is a possibility for Cooper pairs to tunnel between superconductor leads directly. We neglect these modes since they can not propagate alongside the junction and contribute to the PHS. Any increase in the magnitude of $m_X$ makes the width of the Andreev zone narrower. Parts (a) and (b) of Fig.(\ref{Eq.E-phase}) illustrate the energy-phase dispersion of the Andreev modes for two different values of $m_x$ whereas $m_y=0$. The non-symmetric nature of the EPR around $\theta_e=0$,  leads to an imbalance between the upward and downward propagation of Andreev modes and creates the PHS.

As it is obvious from Eq.(\ref{Eq.E-phase}), the parallel component of in-plane magnetization, $m_y \neq 0$, makes an anomalous phase shift on the superconducting phase difference via the term $2m_yd$. Due to the completeness of the Andreev zone, this effect can not create an imbalance between upward and downward modes, and the junction has no PHS. The EPR moves vertically in parts (c) and (d) of Fig(\ref{Fig1.Energy-phase}). This case is known as $\delta\phi$-junction, where the supercurrent has a non-zero value without the superconducting phase difference. 

\subsection{The $\delta\phi$-junction}
\begin{figure}
\includegraphics*[scale=0.35]{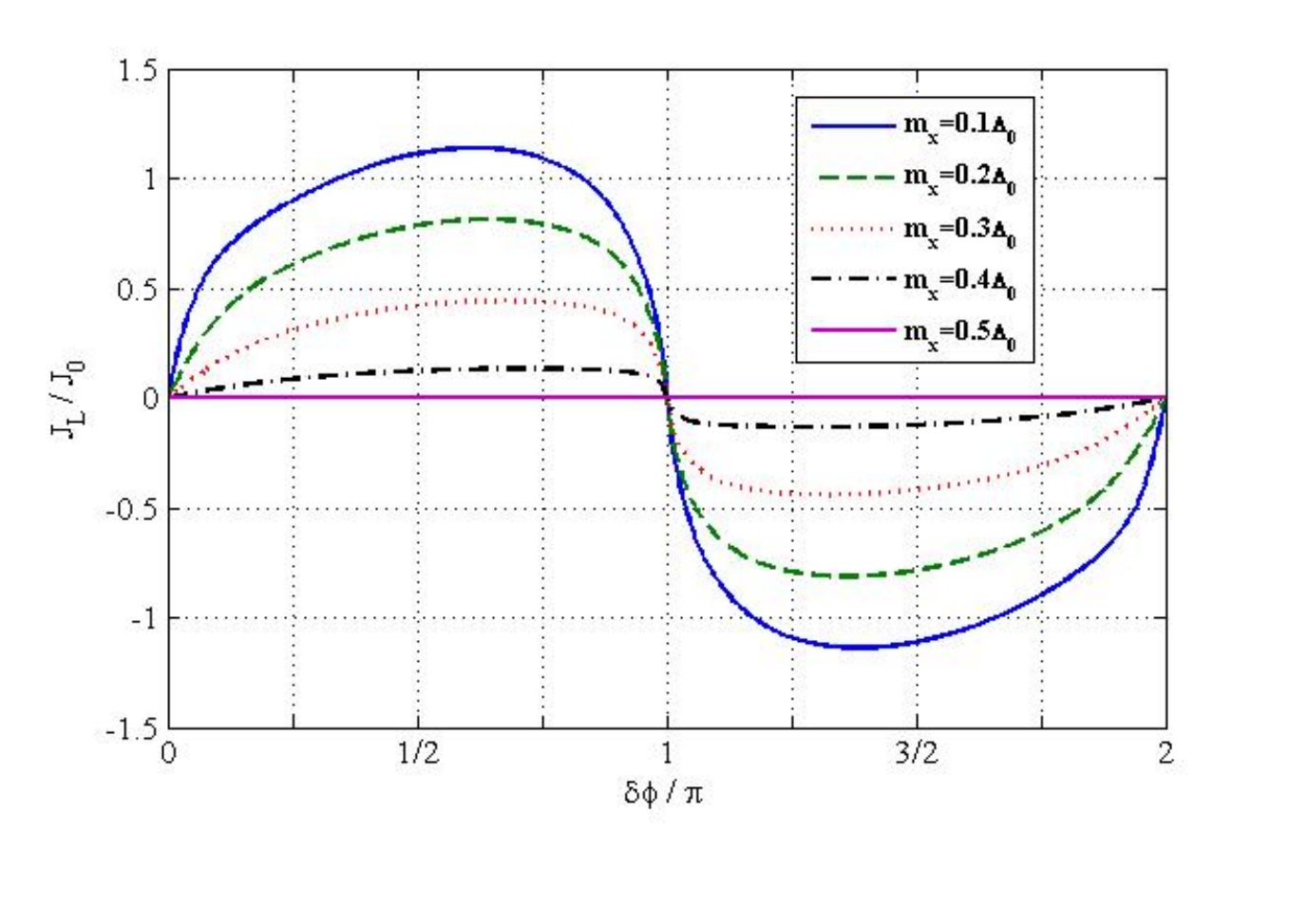}
\caption{(Color online) The Josephson current versus the phase-difference of the junction. Based on the different values of $m_x$. This shows the reduction of critical values accompanied by the $m_x$ increment.  The width of the junction is $d=\hbar v_F/\Delta_0$.}
\label{Fig1.Josephson Current}
\end{figure}

Using the EPR of Eq.(\ref{Eq.E-phase}), the Josephson current in zero temperature that flows between superconducting leads can be obtained via \cite{Titov2006PRB,Linder2010PRL},
\begin{equation}
    J_S(\delta\phi)=\frac{-e}{\hbar}\int_{\frac{-\pi}{2}}^{\frac{\pi}{2}} d\theta_e \frac{\partial\epsilon}{\partial\delta\phi}\left(\cos\theta_e+\cos\theta_h\right).
    \label{Eq.JosephsonCurrent}
\end{equation}
The Josephson current is plotted in Fig(\ref{Fig1.Josephson Current}) for different values of $m_x$. As an apparent effect, any increase in the magnitude of $m_x$, reduces the critical value of supercurrent. This reduction means the $m_x$ bends the direction of the supercurrent to flow parallel to the superconductor interfaces. In the values of $m_x\geq \mu$, those part of the supercurrent that originates the stable part of Andreev modes diminishes, and the junction does not encounter the $\pi$ shift \cite{linder2008PRL}. Its physics can be understood via part (b) of Fig.(\ref{Fig1.ToyModel}), where the two Fermi circles have no overlap for $m_x\geq \mu$ in the short-junction regime.

\begin{figure}
\includegraphics*[scale=0.35]{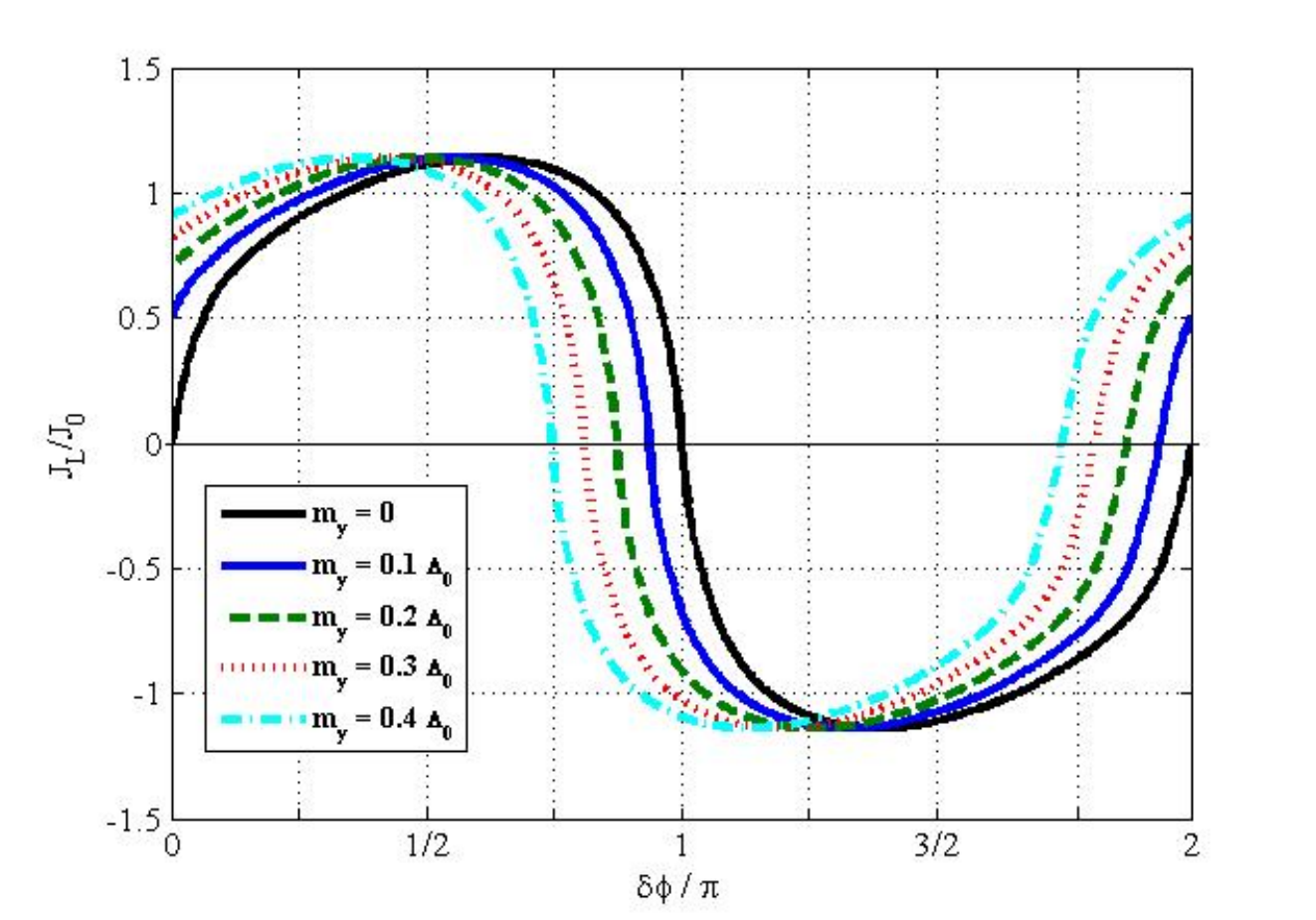}
\caption{(Color online) The Josephson current versus the phase-difference of the junction. In the presence of $m_y$, the current-phase relation moves versus $\delta\phi$ and shows $\delta\phi$-junction. This demonstrates a non-zero supercurrent in the absence of superconducting phase difference. The width of the junction is $d= \hbar v_F/\Delta_0$. }
\label{Fig1.phiCurrent}
\end{figure}

On the other hand, the $m_y$ separates the Fermi circles in $k_x$-direction. This effect is shown by adding an extra term, $2m_yd$, to the superconducting phase difference in Eq.(\ref{Eq.E-phase}). So, the magnitude of $m_y$ and the width of F region lead to the $\delta\phi$-junction,
\begin{equation}
    J_S(\delta\phi)\sim J_C \sin(\delta\phi-2m_yd),
\end{equation}
where $J_C$ is the critical current.
A non-zero supercurrent flows across the junction even without superconducting phase difference. This effect is shown in Fig.(\ref{Fig1.phiCurrent}) for different values of $m_y$.
For $\pi/2d \leq m_y \leq \pi/d$, the junction encounters the $\pi$-shift, and the direction of supercurrent flows in opposite to the superconducting phase difference.

\subsection{The Planar Hall supercurrent}

\begin{figure}
\includegraphics*[scale=0.4]{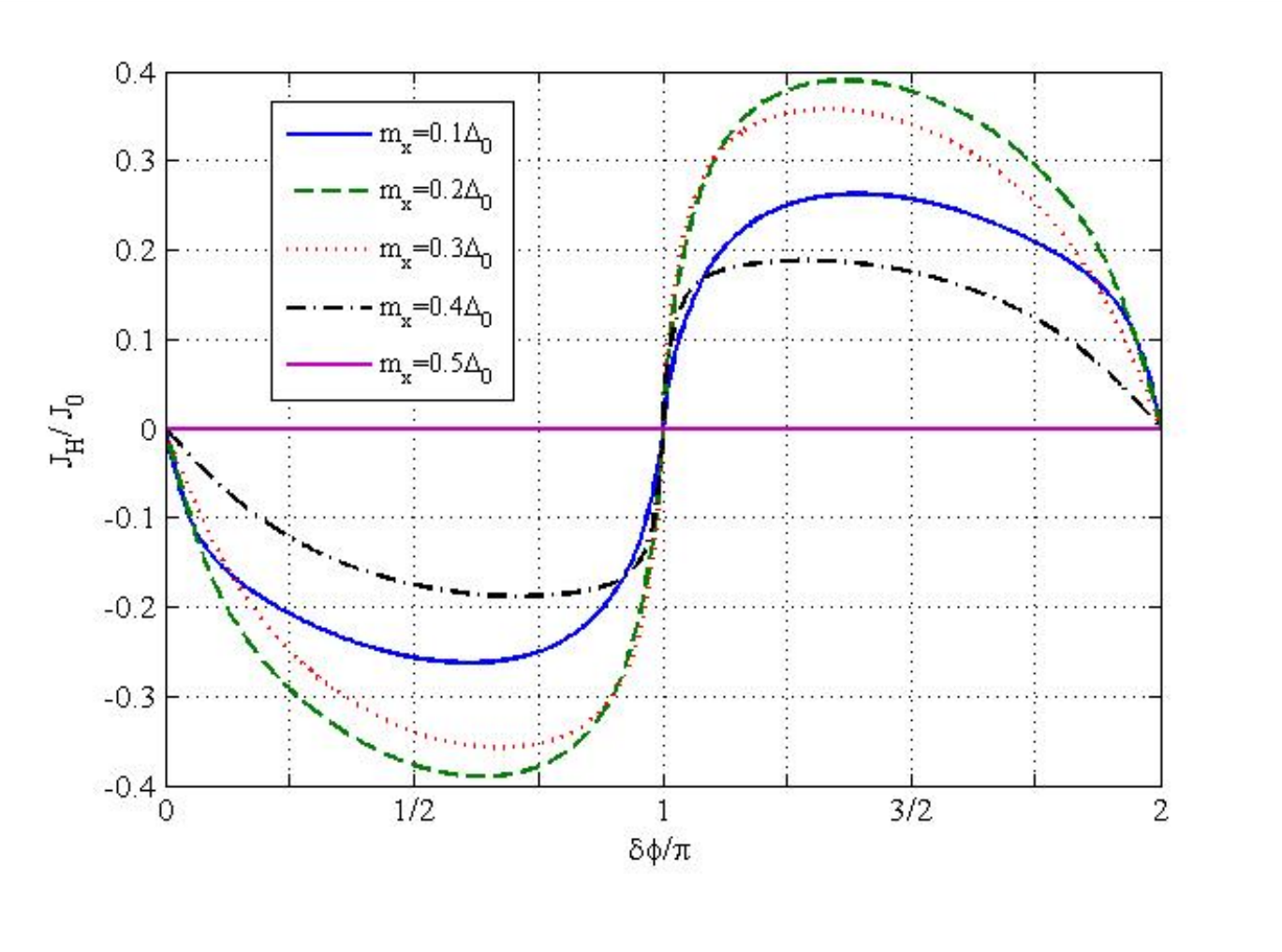}
\caption{(Color online) The Josephson-Hall current versus the superconducting phase-difference of the junction. Based on the different values of $m_x$, it can flow parallel to the junction interface and be detected in real experiments via a four-terminal setup.  The width of the junction is $d=\hbar v_F/\Delta_0$.}
\label{Fig1.HallCurrent}
\end{figure} 

As shown schematically in part(c) of Fig.(\ref{Fig1.ToyModel}), the Andreev modes propagate alongside the junction in the presence of $m_x$. Because of the asymmetric nature of the energy-phase relation, there is an imbalance between upward and downward currents. We modify the previous results to derive the planar Hall supercurrent as below\cite{Titov2007PRB},
\begin{equation}
    J_H=\frac{-e}{\hbar}\int d\theta_e \frac{\partial\epsilon}{\partial\delta\phi}\left(\sin\theta_e+\mathcal{S}\sin\theta_h\right)
    \label{Eq.PlanarHallCurrent}
\end{equation}
The $m_y$ does not affect the planar Hall supercurrent, whereas the $m_x$ directly impacts it. We encounter retro-reflection for Andreev backscattering in the short-junction approximation and non-zero chemical potential, $\mathcal{S}=-1$. Growing $m_x$ creates planar Hall supercurrent until $m_x\sim \mu/2$, where the imbalance between the upward and downward propagation reaches its maximum. Far above this limit, $m_x \geq \mu/2$, the planar Hall supercurrent tends to zero, and it disappears in $m_x = \mu$. Since the components of in-plane magnetization couple to the spin of quasi-particles, Eq.(\ref{Eq.HBdG}), their magnitude and direction move the Dirac cones in $ k$ space. Any change in the spin texture modifies the propagation direction of carriers because of the strong spin-orbit interaction on the surface of topological insulators. The direction of $m_x$ can tune the direction of the planar Hall supercurrent. From the experimental point of view, this provides a suitable tool to design future technology such as spintronics circuits.

\subsection{The spin-transfer torque}

\begin{figure}
\includegraphics*[scale=0.4]{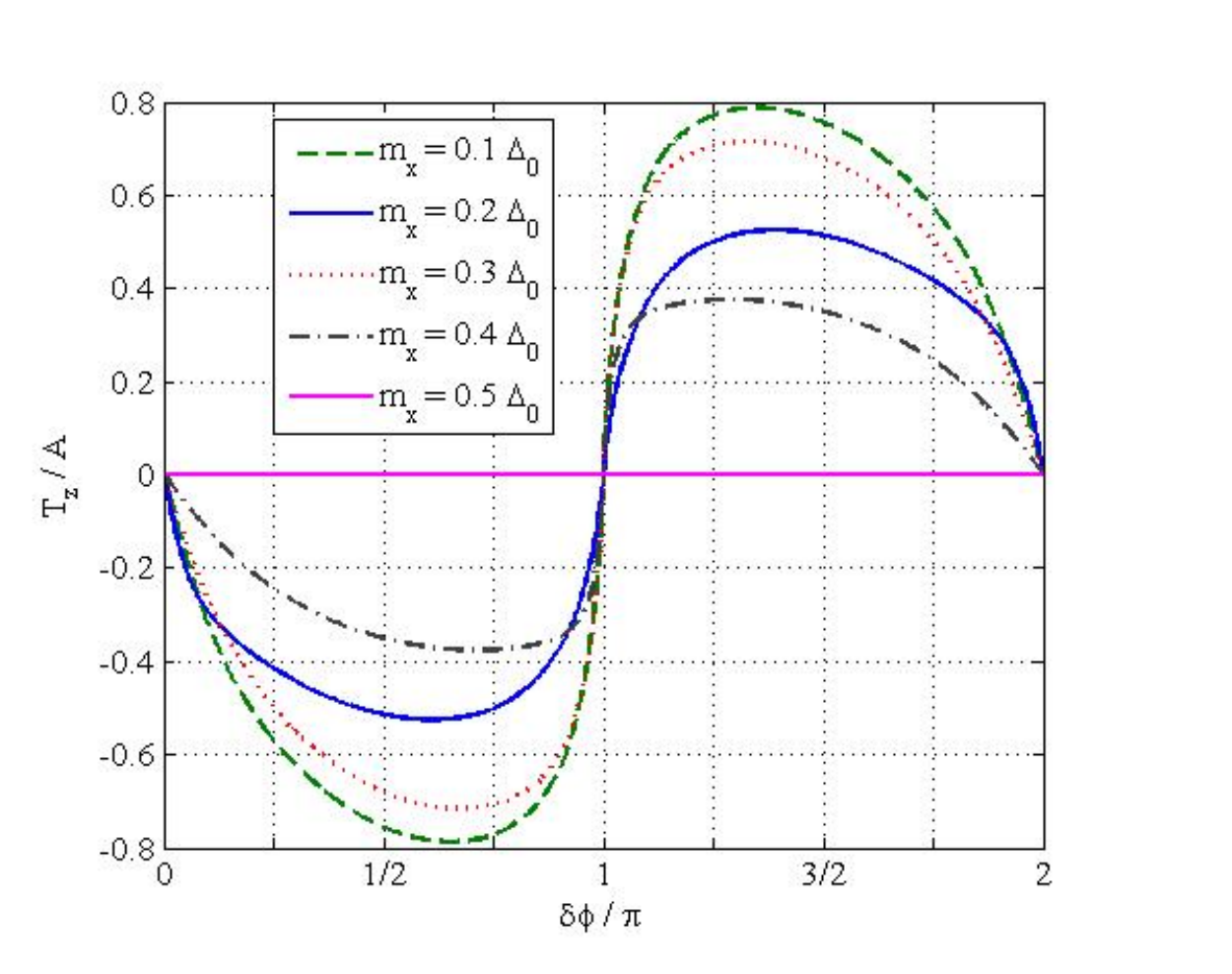}
\caption{(Color online) The spin-transfer torque versus the superconducting phase difference. The magnitude and direction of STT are related to the PHS.  The width of the junction is $d=\hbar v_F/\Delta_0$. }
\label{Fig1.CTT}
\end{figure}

In Eq.(\ref{Eq.CTT2}), we determine the magnitude of the STT imposed on the junction by bending the supercurrent in the presence of $m_x$. Due to the 2D nature of the system, only the $z$-component of the STT is non-zero. This can be derived by considering a square with a length $A$ along the junction.

\begin{equation}
T_z=\oint (J_y dx-J_x dy)=A ( \delta J_y-\delta J_x) 
\label{Eq.CTT3}
\end{equation}

Here, $\delta J_y=J_H$ represents the difference in current passing through the length of the square in the $y$-direction, while $\delta J_x$ denotes the reduction in the Josephson supercurrent passing between the superconducting leads in the $x$-direction. As current is a conserved quantity, the reduction in the $x$-direction corresponds to the phase-coherent supercurrent, $\delta J_x=-J_H$. Consequently, we arrive at

\begin{equation}
T_z \sim 2 J_H A.
\label{Eq.CTTFinal}
\end{equation}

Here, $T_z$ is directly related to the phase-coherent supercurrent. The absence of the $\sigma_z$ term in the spin-orbit coupling term of Eq.(\ref{Eq.HBdG}), along with the two-dimensional nature of the junction, leads to the other components of the STT, $\{T_x,T_y\}$, being zero. The STT imposed on the junction aligns with the direction of the phase-coherent supercurrent. In the range where the planar Hall supercurrent is at its maximum, i.e., $m_x\sim \mu/2$, the STT also reaches its maximum. This arises from the bending of the propagation direction of the Andreev modes. These effects in the energy range of $m_x  \leq \epsilon \leq  \mu/2$ are significant and experimentally detectable. Furthermore, the STT approaches zero at high values of $\mu$ where the displacement of the Dirac cone can be neglected.

\section{Conclusion}
We investigate the influence of in-plane magnetization on a Josephson junction utilizing 3D topological insulators (3DTIs). By employing the Bogoliubov-de Gennes (BdG) formalism under ballistic conditions, we theoretically derive, for the first time, the energy-phase relation (EPR) of Andreev modes. Our analysis reveals that when the in-plane magnetization includes a component perpendicular to the junction's interface, the electron-like and hole-like cones separate in the momentum ($k$) space. Specifically, the $m_x$ component of the magnetization induces an Andreev zone, leading to the propagation of Andreev modes parallel to the junction's interface. Furthermore, the imbalance between the upwards and downwards Andreev modes gives rise to the PHS, accompanied by a reduction in the critical values of the Josephson current flowing between the two superconducting leads. Additionally, we demonstrate how the $m_y$ component adds a phase shift to the superconducting phase difference, creating a $\delta\phi$-junction. Moreover, within a specific range of $m_y$, the junction undergoes a $\pi$-shift. Finally, the PHS imposes a torque on the junction, particularly in the energy range near the Dirac point. The non-zero $z$-component of this torque is correlated with the sign and magnitude of the PHS. Given the importance of spin-transfer torque (STT) in data storage technology, we believe its extension to Josephson junctions could offer valuable insights for the design and fabrication of new devices for a variety of applications.

\bibliography{JosephsonHall}

\end{document}